\documentclass{elsart}
\usepackage{ifpdf}
\usepackage{graphicx,amssymb,lineno}
\ifpdf
\usepackage[%
  pdftitle={Instructions for use of the document class
    elsart},%
  pdfauthor={Simon Pepping},%
  pdfsubject={The preprint document class elsart},%
  pdfkeywords={instructions for use, elsart, document class},%
  pdfstartview=FitH,%
  bookmarks=true,%
  bookmarksopen=true,%
  breaklinks=true,%
  colorlinks=true,%
  linkcolor=blue,anchorcolor=blue,%
  citecolor=blue,filecolor=blue,%
  menucolor=blue,pagecolor=blue,%
  urlcolor=blue]{hyperref}
\else
\usepackage[%
  breaklinks=true,%
  colorlinks=true,%
  linkcolor=blue,anchorcolor=blue,%
  citecolor=blue,filecolor=blue,%
  menucolor=blue,pagecolor=blue,%
  urlcolor=blue]{hyperref}
\fi

\newcommand{\bq}{\begin{equation}}
\newcommand{\eq}{\end{equation}}
\newcommand{\bqa}{\begin{eqnarray}}
\newcommand{\eqa}{\end{eqnarray}}
\newcommand{\nn}{\nonumber \\}

\def\be     {\begin{equation}}
\def\ee     {\end{equation}}
\def\bea        {\begin{eqnarray}}
\def\eea        {\end{eqnarray}}
\def\bnn    {\begin{eqnarray*}}
\def\enn    {\end{eqnarray*}}

\makeatletter
\def\elsartstyle{%
    \def\normalsize{\@setfontsize\normalsize\@xiipt{14.5}}
    \def\small{\@setfontsize\small\@xipt{13.6}}
    \let\footnotesize=\small
    \def\large{\@setfontsize\large\@xivpt{18}}
    \def\Large{\@setfontsize\Large\@xviipt{22}}
    \skip\@mpfootins = 18\p@ \@plus 2\p@
    \normalsize
} \@ifundefined{square}{}{} \makeatother

\begin{document}

\begin{frontmatter}
\title{Quantum phase transition in one dimensional extended Kondo lattice model away from half filling}

\author{Ki-Seok Kim}
\address{School of Physics, Korea Institute for Advanced
Study, Seoul 130-012, Korea}

\ead{kimks@kias.re.kr}

\begin{abstract}
We study one dimensional {\it extended} Kondo lattice model,
described by the $t-J$ Hamiltonian for conduction electrons away
from half filling and the Heisenberg Hamiltonian for localized
spins at half filling. Following Shankar,\cite{Shankar} we find an
effective field theory for this model, where doped holes are
represented by massless Dirac fermions (holons) and spin
excitations are fractionalized into relativistic bosons (spinons).
These holons and spinons interact via U(1) gauge fluctuations.
Effects of Berry phase to the localized spins disappear due to the
presence of Kondo couplings, causing the spinon excitations
gapped. Furthermore, the gauge fluctuations are suppressed by hole
doping. As a result, massive spinons are deconfined to arise in
the localized spins unless the Kondo hybridization is strong
enough. When the Kondo hybridization strength exceeds a certain
value, we find that the localized spin chain becomes critical.
This indicates that the present one dimensional Kondo lattice
model exhibits a phase transition from a spin-gapped phase to a
critical state in the localized spin chain, driven by the Kondo
interaction.
\end{abstract}

\end{frontmatter}

One dimensional Kondo lattice model has been studied intensively.
It seems to be clearly established that in the case of the
half-filled conduction band the ground state is the Kondo
insulator for any non-zero Kondo coupling, which has a gap in both
spin and charge excitations.\cite{Rev_Kondo} Away from half
filling, a paramagnetic metallic phase is expected to
arise.\cite{Rev_Kondo,TTL,LEL} However, its nature remains
controversial. Some numerical and analytical studies support the
Tomonaga-Luttinger liquid with dominant correlations determined by
conduction electrons.\cite{Rev_Kondo,TTL} On the other hand, the
existence of spin gap was reported in the paramagnetic metallic
phase\cite{Rev_Kondo,LEL}.

In the present paper we investigate one dimensional {\it extended}
Kondo lattice model, described by the $t-J$ Hamiltonian for
conduction electrons away from half filling and the Heisenberg
Hamiltonian for localized spins at half filling [Eq. (1)]. An
important difference from the previous studies is that we consider
strongly correlated conduction electrons described by the $t-J$
model instead of non-interacting conduction electrons. Following
Shankar,\cite{Shankar} one can represent the one dimensional $t-J$
model in terms of bosonic spinons for spin-fractionalized
excitations and fermionic holons for doped holes, interacting via
U(1) gauge fluctuations [Eq. (2)]. Spinons carry the spin quantum
number $1/2$ without the charge quantum number while holons carry
the charge quantum number $+e$ without the spin quantum number. On
the other hand, the Heisenberg model for the localized spins can
be represented by bosonic spinons interacting via U(1) gauge
fluctuations with Berry phase [Eq. (2)]. Kondo couplings between
the "conduction" spinons and "localized" spinons are taken into
account [Eq. (2)].

Based on this low energy effective Lagrangian Eq. (2), we study
effects of the Kondo interactions on the fate of spinons. We find
that gapped spinon excitations are deconfined to appear in the
localized spin chain unless the Kondo hybridization is strong
enough. This originates from the fact that the contribution of
Berry phase to the localized spinons disappears due to the Kondo
couplings, causing the spinon excitations gapped, and gauge
fluctuations are suppressed by hole doping, allowing deconfinement
of the gapped spinons. On the other hand, if the Kondo
hybridization strength exceeds a certain value, we find that the
spinon excitations in the localized spin chain become critical.
This indicates that the present one dimensional Kondo lattice
model exhibits a phase transition from a spin-gapped phase to a
critical state, driven by the Kondo interaction.

We consider a hole-doped antiferromagnetic spin chain in the
presence of Kondo couplings with half-filled local magnetic
moments \bqa && H = H_{c} + H_{m} + H_{K} , \nn && H_{c} =
-t\sum_{i=1}^{N}(c_{\sigma{i}}^{\dagger}c_{\sigma{i+1}} + h.c.) +
J\sum_{i=1}^{N}{\bf s}_{i}\cdot{\bf s}_{i+1} , \nn && H_{m} =
I\sum_{i=1}^{N}{\bf \tau}_{i}\cdot{\bf \tau}_{i+1} , \nn && H_{K}
= J_{K}\sum_{i=1}^{N}{\bf s}_{i}\cdot{\bf \tau}_{i} . \eqa Here
the $t-J$ Hamiltonian $H_{c}$ describes a hole-doped
antiferromagnetic spin chain corresponding to the conduction band
in the Kondo lattice model, where $t$ and $J$ represent the
strength of hopping and antiferromagnetic correlations for the
conduction electrons, respectively. The Heisenberg Hamiltonian
$H_{m}$ depicts a half-filled spin chain of localized magnetic
moments, where $I$ is the strength of antiferromagnetic
correlations between the localized spins. Strongly correlated
conduction electrons and localized magnetic moments are
antiferromagnetically correlated via the Kondo coupling $H_{K}$
with the strength $J_K$. If strong electron-electron correlations
represented by the Heisenberg coupling term $J\sum_{i=1}^{N}{\bf
s}_{i}\cdot{\bf s}_{i+1}$ and the no-double-occupancy constraint
$\sum_{\sigma=1}^{2}c_{\sigma{i}}^{\dagger}c_{\sigma{i}} \le 1$
are neglected, this Hamiltonian is naturally reduced to the
conventional Kondo lattice model. In this respect we deal with
both electron-electron and electron-local moment interactions on
an equal footing in the presence of hole doping.

In passing, we review physics of the one dimensional $t-J$
Hamiltonian. In the absence of hole doping hopping of electrons is
suppressed, thus the $t-J$ Hamiltonian is reduced to the
Heisenberg Hamiltonian describing the antiferromagnetic spin
chain. Low energy physics of the quantum spin chain can be
described by the O(3) non-linear $\sigma$ model with Berry phase.
Utilizing the $CP^1$ representation, one can express the
non-linear $\sigma$ model in terms of bosonic spinons interacting
via compact U(1) gauge fields in the presence of the Berry phase
contribution. Since the Berry phase term is ignorable in the case
of integer spin, strong quantum fluctuations originating from low
dimensionality lead the integer spin chain to be disordered,
causing the bosonic spinons gapped.\cite{Nagaosa} These massive
spinons are confined via strong gauge fluctuations, resulting in
spin excitons (spinon-antispinon bound states) as elementary
excitations.\cite{Nagaosa,Deconfinement} In the case of half-odd
integer spin the Berry phase plays a crucial role to cause
destructive interference between quantum fluctuations, thus
weakening the quantum fluctuations. Owing to the Berry phase
contribution the half-odd integer spin chain is expected to be
ordered. But, low dimensionality leads the system to be not
ordered but critical, causing the spinons gapless.\cite{Nagaosa}
These massless spinons are deconfined because critical
fluctuations of the spinons weaken gauge fluctuations via
screening.\cite{Deconfinement,Kim_DQCP} In this respect the
half-odd integer spin chain is considered to be the
Tomonaga-Luttinger liquid.\cite{Nagaosa}

When holes are doped to the antiferromagnetic spin chain, Shankar
showed that doped holes can be expressed by massless Dirac
fermions dubbed holons and these fermionic holons interact with
the bosonic spinons via U(1) gauge fluctuations.\cite{Shankar} The
presence of massless Dirac fermions completely alters the
resulting phase in the absence of those. Massless Dirac fermions
are well known to kill the Berry phase contribution.\cite{Witten}
Then, the bosonic spinons in the doped half-odd integer spin chain
are expected to be massive like those in the undoped integer spin
chain. But, these spinons are not confined in contrast to the case
of the integer spin chain\cite{Witten,Shankar} because the gauge
fluctuations become massive due to the massless Dirac fermions,
thus ignored in the low energy limit.\cite{QED2} A spin liquid
state with gapped spinon excitations emerges in the doped
antiferromagnetic spin chain. Furthermore, one can find that there
are superconducting correlations of doped holes, resulting from
gapless charge fluctuations (holons).\cite{Shankar} It should be
noted that this result is exact in the low energy
limit.\cite{Shankar}

The present problem is more complex than the above owing to the
presence of Kondo couplings with local magnetic moments. Following
Shankar,\cite{Shankar} we obtain an effective Lagrangian for Eq.
(1) \bqa && S_{eff} = S_{c} + S_{m} + S_{K} , \nn && S_{c} =
\int{d^2x} \Bigl[ |(\partial_{\mu} - ia_{\mu})z_{\sigma}|^2 +
m_{z}^2|z_{\sigma}|^2 + \frac{u_z}{2}(|z_{\sigma}|^2)^2 +
\frac{1}{2g_{z}^{2}}|\epsilon_{\mu\nu}\partial_{\mu}a_{\nu}|^{2} -
iS\epsilon_{\mu\nu}\partial_{\mu}a_{\nu} \Bigr] \nn && +
\int{d^2x} \Bigl[
\bar{\psi}_{A}\gamma_{\mu}(\partial_{\mu}+ia_{\mu})\psi_{A} +
\bar{\psi}_{B}\gamma_{\mu}(\partial_{\mu}-ia_{\mu})\psi_{B} \Bigr]
, \nn && S_{m} = \int{d^2x} \Bigl[ |(\partial_{\mu} -
ic_{\mu})b_{\sigma}|^2 + m_{b}^2|b_{\sigma}|^2 +
\frac{u_b}{2}(|b_{\sigma}|^2)^2 +
\frac{1}{2g_{b}^{2}}|\epsilon_{\mu\nu}\partial_{\mu}c_{\nu}|^{2} -
iS\epsilon_{\mu\nu}\partial_{\mu}c_{\nu} \Bigr] , \nn && S_{K} =
\int{d^2x}
\frac{J_K}{4}z_{\alpha}^{\dagger}\sigma^{l}_{\alpha\beta}z_{\beta}\cdot
b_{\gamma}^{\dagger}\sigma^{l}_{\gamma\delta}b_{\delta} . \eqa
Here $z_{\sigma}$ and $\psi_{A(B)}$ represent a bosonic spinon
with spin $\sigma = \uparrow, \downarrow$ and a fermionic holon in
a sublattice $A(B)$ in the conduction chain, respectively. The
spinons and holons interact via the U(1) gauge field $a_{\mu}$
with the coupling strength $g_z$. $m_z$ and $u_z$ are the mass and
local interaction strength of the conduction spinons,
respectively. $S$ in the Berry phase term
$iS\epsilon_{\mu\nu}\partial_{\mu}a_{\nu}$ represents the value of
spin $1/2$. $b_{\sigma}$ is a bosonic spinon in the chain of local
magnetic moments. These spinons interact with each other via the
other U(1) gauge field $c_{\mu}$ with the coupling strength $g_b$.
$m_b$ and $u_b$ are the mass and local interaction strength of the
local spinons, respectively.

New physics would arise from the Kondo coupling term $S_{K}$
between the two spinons, $z_\sigma$ and $b_{\sigma}$. In order to
treat the Kondo coupling term, we perform the usual
Hubbard-Stratonovich transformation, and obtain the following one
body effective action\cite{Kondo1,Kondo2} $S_{K} = \int{d^2x}
\Bigl[ \frac{4}{J_K}\Delta_{0}^{2} -
\Delta_{0}e^{i\phi}z_{\sigma}^{\dagger}b_{\sigma} - h.c. \Bigr]$.
Here $\Delta_0$ is an amplitude of the hybridization order
parameter $\Delta = \Delta_{0}e^{i\phi}$ between the $z_\sigma$
and $b_\sigma$ spinons, and $\phi$ is its phase-fluctuation field.
The amplitude is given by $\Delta_0 =
\frac{J_{K}}{2}|<z_{\sigma}^{\dagger}b_{\sigma}>|$. In the present
paper we concentrate on phase fluctuations of the hybridization
order parameter. They are controlled by the low energy effective
action $S_{\phi} = \int {d^2x} \frac{\rho}{2}|\partial_{\mu}\phi -
a_{\mu} + c_{\mu}|^{2}$, where $\rho$ is the stiffness parameter
proportional to $\Delta_{0}^{2}$.\cite{Kondo2}

In order to take phase fluctuations in the Kondo coupling term, we
perform the gauge transformation of $\tilde{b}_{\sigma} =
e^{i\phi}b_{\sigma}$ and $\tilde{c}_{\mu} = c_{\mu} +
\partial_{\mu}\phi$.\cite{Kondo2} Then, Eq. (2) reads
\bqa && S_{eff} = \int{d^2x} \Bigl[ |(\partial_{\mu} -
i{a}_{\mu}){z}_{\sigma}|^2 + m_{z}^2|{z}_{\sigma}|^2 +
\frac{u_z}{2}(|{z}_{\sigma}|^2)^2 +
\frac{1}{2g_{z}^{2}}|\epsilon_{\mu\nu}\partial_{\mu}a_{\nu}|^{2} -
iS\epsilon_{\mu\nu}\partial_{\mu}a_{\nu} \Bigr] \nn && +
\int{d^2x} \Bigl[
\bar{\psi}_{A}\gamma_{\mu}(\partial_{\mu}+ia_{\mu})\psi_{A} +
\bar{\psi}_{B}\gamma_{\mu}(\partial_{\mu}-ia_{\mu})\psi_{B} \Bigr]
\nn && + \int{d^2x} \Bigl[ |(\partial_{\mu} -
i\tilde{c}_{\mu})\tilde{b}_{\sigma}|^2 +
m_{b}^2|\tilde{b}_{\sigma}|^2 +
\frac{u_b}{2}(|\tilde{b}_{\sigma}|^2)^2 \nn && +
\frac{1}{2g_{b}^{2}}|\epsilon_{\mu\nu}\partial_{\mu}\tilde{c}_{\nu}
- \epsilon_{\mu\nu}\partial_{\mu}\partial_{\nu}\phi |^{2} -
iS\epsilon_{\mu\nu}\partial_{\mu}\tilde{c}_{\nu} +
iS\epsilon_{\mu\nu}\partial_{\mu}\partial_{\nu}\phi \Bigr] \nn &&
+ \int{d^2x}\Bigl[\frac{\rho}{2}|{a}_{\mu} - \tilde{c}_{\mu}|^{2}
- \Delta_{0}{z}_{\sigma}^{\dagger}\tilde{b}_{\sigma} - h.c. +
\frac{4}{J_K}\Delta_{0}^{2}\Bigr] . \eqa In the action of the
renormalized local spinons $\tilde{b}_{\sigma}$ the Berry
phase-induced term $iS\int{d^2x}
\epsilon_{\mu\nu}\partial_{\mu}\partial_{\nu}\phi$ can be
expressed by $iS\int{d^2x}
\epsilon_{\mu\nu}\partial_{\mu}\partial_{\nu}\phi =
iS\int{dx_{\mu}}\partial_{\mu}\phi = i2\pi{S}n = i\pi{n}$. This is
nothing but a vortex contribution in the phase field $\phi$. $n$
is an integer representing a vortex charge. Vortex excitations
would break phase coherence, making the hybridization disappear.
But, the Berry phase-induced term plays a role to suppress single
vortex excitations. When a single vortex configuration appears,
this complex phase term gives a factor $-1$ to the partition
function of the action Eq. (3). This destructive interference
originating from the Berry phase contribution for single vortex
excitations leads to suppression of single vortex excitations,
indicating no vortex condensation. As a result, phase coherence is
sustained, i.e., $<e^{i\phi}> \not= 0$, resulting in the
hybridization of the two spinons.

This mechanism is similar to that for the suppression of strong
gauge fluctuations (instantons) between topologically inequivalent
gauge vacua owing to the Berry phase contribution in the half-odd
integer spin chain,\cite{Deconfinement} as discussed before. The
suppression of instanton excitations in gauge fluctuations
increases the tendency of antiferromagnetic ordering. In a
different angle the condensation of the order parameter $\Delta$
means that the Kondo couplings are relevant in the context of
renormalization group. The relevance of the Kondo couplings causes
the mass term $\frac{\rho}{2}|{a}_{\mu} - \tilde{c}_{\mu}|^{2}$.
This is nothing but the Anderson-Higgs mechanism, allowing us to
set $\tilde{c}_{\mu} = a_{\mu}$ in the low energy limit. Then, we
obtain the following effective action \bqa && S_{eff} = \int{d^2x}
\Bigl[ |(\partial_{\mu} - i{a}_{\mu}){z}_{\sigma}|^2 +
m_{z}^2|{z}_{\sigma}|^2 + \frac{u_z}{2}(|{z}_{\sigma}|^2)^2 +
\frac{1}{2g_{z}^{2}}|\epsilon_{\mu\nu}\partial_{\mu}a_{\nu}|^{2}
\Bigr] \nn && + \int{d^2x} \Bigl[
\bar{\psi}_{A}\gamma_{\mu}(\partial_{\mu}+ia_{\mu})\psi_{A} +
\bar{\psi}_{B}\gamma_{\mu}(\partial_{\mu}-ia_{\mu})\psi_{B} \Bigr]
\nn && + \int{d^2x} \Bigl[ |(\partial_{\mu} -
ia_{\mu})\tilde{b}_{\sigma}|^2 + m_{b}^2|\tilde{b}_{\sigma}|^2 +
\frac{u_b}{2}(|\tilde{b}_{\sigma}|^2)^2 +
\frac{1}{2g_{b}^{2}}|\epsilon_{\mu\nu}\partial_{\mu}a_{\nu}|^{2}
\Bigr] \nn && + \int{d^2x}\Bigl[\frac{4}{J_K}\Delta_{0}^{2} -
\Delta_{0}{z}_{\sigma}^{\dagger}\tilde{b}_{\sigma} - h.c. \Bigr] .
\eqa Surprisingly, only one gauge fluctuations remain as a result
of the relevant Kondo coupling. A similar result is also obtained
in Ref. \cite{Kondo2}. Furthermore, considering $2S = 1$, one can
easily find that the following result
$iS\epsilon_{\mu\nu}\partial_{\mu}a_{\nu} +
iS\epsilon_{\mu\nu}\partial_{\mu}\tilde{c}_{\nu} =
i2S\epsilon_{\mu\nu}\partial_{\mu}a_{\nu}$ yields the Berry phase
contribution to vanish. This is well known in the non-linear
$\sigma$ model approach to the spin-ladder system.\cite{Ladder}

We check whether our present analysis reproduces the known results
in the absence of hole doping. Considering a half-filled
conduction chain, the present problem is the same as the
two-leg-ladder problem. It is well known that the ground state is
the phase of Kondo singlets.\cite{Schulz} Elementary excitations
are singlet to triplet excitations with spin $1$.\cite{Schulz} Eq.
(4) without massless Dirac fermions recovers these well known
properties. As discussed earlier, Eq. (4) shows condensation of
the hybridization order parameter $<\Delta> \not= 0$, indicating
the presence of Kondo singlets. The formation of Kondo singlets
gives a mass gap to both the $z_\sigma$ and $\tilde{b}_{\sigma}$
spinons. These massive spinons are confined via strong gauge
fluctuations $a_{\mu}$.\cite{Nagaosa} As a result, elementary
excitations are Kondo triplet excitations
$z^{\dagger}_{\sigma}\tilde{b}_{\sigma}$ with spin $1$.

Now we investigate the role of hole doping in this system. In
order to treat massless Dirac fermions, we utilize the well known
bosonization technique\cite{Shankar,Witten} \bqa &&
\bar{\psi}_{A}\gamma_{\mu}\partial_{\mu}\psi_{A} =
\frac{1}{2}(\partial_{\mu}\theta_{A})^2 \mbox{,  }  ~~~~~
\bar{\psi}_{B}\gamma_{\mu}\partial_{\mu}\psi_{B} =
\frac{1}{2}(\partial_{\mu}\theta_{B})^2 , \nn &&
\bar{\psi}_{A}\gamma_{\mu}\psi_{A} =
\frac{1}{\sqrt{\pi}}\epsilon_{\mu\nu}\partial_{\nu}\theta_{A}
\mbox{, }  ~~~~~ \bar{\psi}_{B}\gamma_{\mu}\psi_{B} =
\frac{1}{\sqrt{\pi}}\epsilon_{\mu\nu}\partial_{\nu}\theta_{B} ,
\eqa where $\theta_{A}$ and $\theta_{B}$ are bosonic fields in
each sublattice. These bosonic field variables are associated with
collective density fluctuations of the Dirac fermions.

Inserting these into the above action Eq. (4), we obtain an
effective action \bqa && S_{eff} = \int{d^2x} \Bigl[
|(\partial_{\mu} - ia_{\mu}){z}_{\sigma}|^2 +
m_{z}^2|{z}_{\sigma}|^2 + \frac{u_z}{2}(|{z}_{\sigma}|^2)^2 +
\frac{1}{2g_{z}^{2}}|\epsilon_{\mu\nu}\partial_{\mu}a_{\nu}|^{2}
\Bigr] \nn && + \int{d^2x} \Bigl[
\frac{1}{2}(\partial_{\mu}\theta_{+})^2 +
\frac{1}{2}(\partial_{\mu}{\theta}_{-})^2 +
i\sqrt{\frac{2}{\pi}}{\theta}_{-}\epsilon_{\mu\nu}\partial_{\mu}a_{\nu}\Bigr]
\nn && + \int{d^2x} \Bigl[ |(\partial_{\mu} -
ia_{\mu})\tilde{b}_{\sigma}|^2 + m_{b}^2|\tilde{b}_{\sigma}|^2 +
\frac{u_b}{2}(|\tilde{b}_{\sigma}|^2)^2 +
\frac{1}{2g_{b}^{2}}|\epsilon_{\mu\nu}\partial_{\mu}a_{\nu}|^{2}
\Bigr] \nn && + \int{d^2x}\Bigl[\frac{4}{J_K}\Delta_{0}^{2} -
\Delta_{0}{z}_{\sigma}^{\dagger}\tilde{b}_{\sigma} - h.c. \Bigr]
\eqa with $\theta_{+} = \frac{1}{\sqrt{2}}(\theta_{A} +
\theta_{B})$ and $\theta_{-} = \frac{1}{\sqrt{2}}(\theta_{A} -
\theta_{B})$. The presence of massless Dirac fermions results in a
mass gap to the U(1) gauge field
$a_{\mu}$.\cite{Witten,Shankar,QED2} Integrating over the
$\theta_{-}$ fields, one can see that a mass term of the gauge
field appears, allowing to ignore the U(1) gauge fluctuations in
the low energy limit. This causes deconfinement of the massive
spinons in contrast to the half-filled case.

The resulting low energy effective action is given by \bqa &&
S_{eff} = \int{d^2x} \Bigl[ |\partial_{\mu}z_{\sigma}|^2 +
m_{z}^2|{z}_{\sigma}|^2 + |\partial_{\mu}\tilde{b}_{\sigma}|^2 +
m_{b}^2|\tilde{b}_{\sigma}|^2 + \frac{4}{J_K}\Delta_{0}^{2} -
\Delta_{0}{z}_{\sigma}^{\dagger}\tilde{b}_{\sigma} - h.c.\Bigr]
\nn && + \int{d^2x} \frac{1}{2}(\partial_{\mu}\theta_{+})^2 . \eqa
Performing the canonical transformation for the $z_{\sigma}$ and
$\tilde{b}_{\sigma}$ bosons, one can solve the Kondo hybridization
to find Bogliubov quasispinons. The mass gap of the quasispinon
excitation is given by $M_{\pm}^{2} = \frac{1}{2}\Bigl(m_{z}^{2} +
m_{b}^{2} \pm \sqrt{(m_{z}^{2} - m_{b}^{2})^{2} +
(2\Delta_{0})^{2}}\Bigr)$, originating from electron-electron and
electron-local moment interactions. Unless the Kondo hybridization
$\Delta_{0}$ is strong enough to satisfy $\Delta_{0} \geq
m_{z}m_{b}$, both quasispinon excitations are gapped although this
spin gap is exponentially small. This indicates that despite half
filling the massive spinons are deconfined to emerge in the local
spin chain, allowed by the relevant Kondo coupling and hole doping
to the conduction chain. On the other hand, if $\Delta_{0} \geq
m_{z}m_{b}$ is satisfied, low energy quasispinons become condensed
owing to $M_{-}^{2} \leq 0$. Because low dimensionality prohibits
the localized spin chain from ordering, it would be critical. In
this case the Tomonaga-Luttinger liquid is expected. Introducing
an electromagnetic field $A_{\mu}$, we obtain the coupling term of
$i\sqrt{\frac{2}{\pi}}\theta_{+}\epsilon_{\mu\nu}\partial_{\mu}A_{\nu}$.
A mass of the electromagnetic field arises when the $\theta_{+}$
fields are integrated out. This implies superconductivity in the
doped spin chain, consistent with the result of
Shankar.\cite{Shankar}

In this paper we found that the present one dimensional Kondo
lattice model [Eq. (1)] shows a phase transition from a
spin-gapped phase to a spin-critical state as increasing the Kondo
coupling strength. Resorting to the effective field theory [Eq.
(2)] for this model, we showed that Berry phase vanishes due to
Kondo couplings [Eq. (4)] and U(1) gauge fluctuations are
suppressed by hole doping [Eq. (6)]. From the resulting effective
action [Eq. (7)] we found that gapped spinon excitations can
emerge in the half-filled antiferromagnetic spin chain when the
Kondo hybridization is not strong enough, while the spinon
excitations can be critical if the Kondo hybridization strength is
beyond a certain value. We expect that the present formulation can
be easily extended in two spacial
dimensions.\cite{Kim_Kondo_Boson} It will be interesting to
examine the issue of the spinon deconfinement in the two
dimensional Kondo lattice model.


\begin{thebibliography}{9}
\bibitem{Shankar} R. Shankar, Phys. Rev. Lett. {\bf 63}, 203
(1989); R. Shankar, Nucl. phys. B {\bf 330}, 433 (1990).
\bibitem{Rev_Kondo} H. Tsunetsugu, M. Sigrist, and K. Ueda,
Rev. Mod. Phys. {\bf 69}, 809 (1997), and references therein.
\bibitem{TTL} H. Tsunetsugu, M. Sigrist, and K. Ueda, Phys. Rev. B
{\bf 47}, 8345 (1993); S. Moukouri and L. G. Caron, Phys. Rev. B
{\bf 54}, 12212 (1996); N. Shibata, A. Tsvelik, and K. Ueda, Phys.
Rev. B {\bf 56}, 330 (1997); S. Fujimoto and N. Kawakami, J. Phys.
Soc. Jpn. {\bf 63}, 4322 (1994).
\bibitem{LEL} S. R. White and I. Affleck, Phys. Rev. B {\bf 54},
9862 (1996); O. Zachar, S. A. Kivelson, and V. J. Emery, Phys.
Rev. Lett. {\bf 77}, 1342 (1996); Karyn Le Hur, Phys. Rev. B {\bf
58}, 10261 (1998).
\bibitem{Nagaosa} N. Nagaosa, Quantum Field Theory in
Strongly Correlated Electronic Systems (Springer-Verlag, Berlin,
1999).
\bibitem{Deconfinement} T. Senthil, A. Vishwanath, L. Balents, S. Sachdev, and M. P. A.
Fisher, Science {\bf 303}, 1490 (2004); T. Senthil, L. Balents, S.
Sachdev, A. Vishwanath, and M. P.A. Fisher, Phys. Rev. B {\bf 70},
144407 (2004), and references therein.
\bibitem{Kim_DQCP} Ki-Seok Kim, Phys. Rev. B {\bf 72}, 035109
(2005).
\bibitem{Witten} E. Witten, Nucl. phys. B {\bf 149}, 285 (1979).
\bibitem{QED2} J. Zinn-Justin, Quantum Field Theory and Critical
Phenomena (second edition), Oxford University Press (1993); D. H.
Kim and P. A. Lee, Annals Phys. {\bf 272}, 130 (1999), and
references therein.
\bibitem{Kondo1} T. Senthil, M. Vojta, and S. Sachdev,
Phys. Rev. B {\bf 69}, 035111 (2004).
\bibitem{Kondo2} Ki-Seok Kim, Phys. Rev. B {\bf 71}, 205101
(2005).
\bibitem{Ladder} G. Sierra, J. Phys. A {\bf 29}, 3229 (1996).
\bibitem{Schulz} H.J. Schulz, G. Cuniberti, and P. Pieri,
Field Theories for Low-Dimensional Condensed Matter Systems, G.
Morandi et al. Eds. Springer (2000).
\bibitem{Kim_Kondo_Boson} Ki-Seok Kim, Phys. Rev. B {\bf 72},
144426 (2005).
\end{thebibliography}
\end{document}